\documentclass[prb,twocolumn]{revtex4}

\usepackage{graphicx}

\begin{document}

\title{A digital oscilloscope setup for the measurement of a transistor's characteristics}

\author{Pierre de Buyl}
\address{Center for Nonlinear Phenomena and Complex Systems \\
 Universit{\'e} Libre de Bruxelles (U.L.B.), Code Postal 231, Campus Plaine, B-1050 Brussels, Belgium}

\begin{abstract}
  The measure of the characteristics of a transistor is an important step in an introductory electronics course. We propose to use a digital oscilloscope with a USB connection to perform a measurement of the characteristic curves with no additional custom circuitry. The setup is presented alongside with code that allows the importation and analysis of the results with open-source software.
\end{abstract}

\maketitle

\section{Introduction}

Experimental work in a physics curriculum needs a good understanding of physical theory and hopefully helps to achieve to understand the same physics. Mastery of technological tools that allow the control and measurement of physical phenomena play a key role in the experimental process.
The tools range from the most simple (a voltmeter for instance) to specialized equipment. A common tool that is encountered in many laboratories is the oscilloscope, from the first year electricity lab to advanced research labs.
While the oscilloscope allows a direct measurement of many quantities, a non-trivial task is the automated measurement of a transistor's characteristics.\cite{malvino} Dedicated circuit have been devised however to make the process possible \cite{gubanski_tpt_1973,barnes_tpt_1974,ponting_tpt_1991,ramachandran_physed_1993,camden_tpt_2005} as well as commercial curve tracers.
The measurement of characteristics is a basic experiment in introductory electronics courses and we will use it as an example, in association with a digital storage oscilloscope (DSO). Despite the commonness of DSOs, few articles in physics journal are dedicated to its use (see, e.g., Refs.~\onlinecite{manzanares_ajp_1994,masters_ajp_1997,potter_tpt_2003,martinez_ricci_ajp_2007,wadhwa_physed_2009}).

We propose in this article a setup to analyze the data on a computer via inexpensive means~: a USB stick and open-source software. The motivation behind this choice is two-fold~: the cost is negligible once a suitable oscilloscope is available in the lab and the software is available for all the students to perform the analysis on whichever computer they use.

\section{The setup}
\label{sec:setup}

We use a Tektronix TDS-1001B DSO, but any oscilloscope with the ability to write files on a USB device should work, and a function generator.
The setup is made via the ``SAVE/RECALL'' button. While we refer the reader to the manual for the setup, here is what we expect to find after a push on the ``PRINT'' key~: a folder on the USB drive with one CSV file per channel, a setup file and a screenshot of the scope.
The screenshot allows one to visually identify a measure. It is however necessary to take note of the folder name that appears on the screen and to associate it with a given experiment.

Once a file is written to the USB device, one can read it on the computer. The CSV file can be used with common spreadsheet programs for instance, but we propose to use an open-source library that can be readily be installed by the students at no cost. All the file operations, computations and plotting will be made with the help of Python and Matplotlib,\cite{python,mpl} a plotting library. All the code necessary to perform the work presented in this article is available in the supplementary material to this article, as well as the data files.
We illustrate however the core commands issued to obtain the results in order to emphasize the simplicity offered by this solution~:
in a terminal, one needs to go to the directory in which the data files reside, then launch python and load the libraries with (the ``\texttt{>>>}'' represents the Python shell)~:
\begin{verbatim}
>>> from matplotlib.mlab import csv2rec
>>> from pylab import *
>>> ion()
\end{verbatim}
the channels are loaded separately via the commands
\begin{verbatim}
>>> CH1 = csv2rec('F0000CH1.CSV',skiprows=18,
      names=['A','B','C','t','V'])
>>> CH2 = csv2rec('F0000CH2.CSV',skiprows=18,
      names=['A','B','C','t','V'])
\end{verbatim}
where ``0000'' is changed according to the folded in which the data is saved. The procedure must be repeated for each set of measures (more sophisticated instructions are found in the supplementary material).

\section{Measure of the characteristic curves}
\label{sec:char}

\begin{figure}[ht]
  \centering
  \includegraphics[width=.9\linewidth]{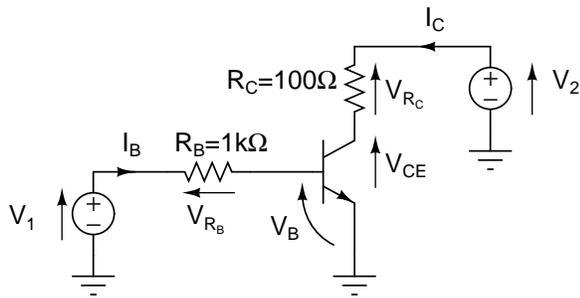}
  \caption{The circuit used to measure the characteristic curves (base-emitter {\it and} collector-emitter) of the transistor. See text for use of the voltage sources $V_1$ and $V_2$.}
  \label{fig:char-circ}
\end{figure}

The measure of the characteristic curves is made thanks to the circuit presented in Fig.~\ref{fig:char-circ}. Of the two voltage sources $V_1$ and $V_2$, one will be set to a constant DC value while the other one will be set to span a voltage range, allowing to trace characteristics curves. In addition to the transistor and the DC and AC voltage sources, the material needed is comprised of one $1$k$\Omega$ resistor and one $100\Omega$ ($1$W) resistor. The transistor under test is here a 2N2219A NPN transistor.

\subsection{The base-emitter characteristic}
\label{sec:BE}

The measurement of the base-emitter (BE) characteristic is based on the voltage across the BE junction of the transistor and on the current $I_B$ flowing through the base.
$I_B$ is measured as the voltage drop across the resistance $R_B$ divided by $R_B$ as the oscilloscope cannot measure directly a current.
We expect a typical current $I_B$ that is of order $1$mA and use a value of $R_B=1$k$\Omega$.

A $12$V voltage source is provided ($V_2$ is set to $12$V) to polarize the collector-emitter junction and the voltage source $V_1$ is a sawtooth signal of frequency $\nu = 100$Hz and a voltage of $0$V to $1$V. A single period of the signal is sufficient to collect the complete characteristic, as it allows a sweep of $V_{BE}$ from $0$ to about $0.75$V. $I_B$ varies between $0$ and $0.3~$mA.

\begin{figure}
  \centering
  \includegraphics[width=.9\linewidth]{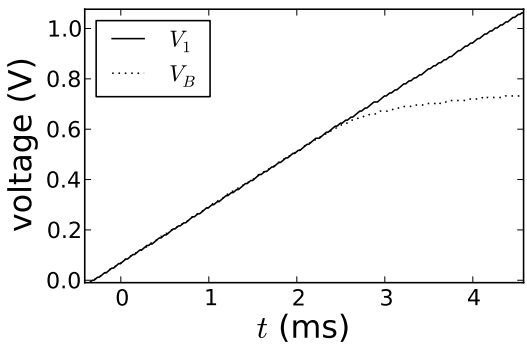}
  \caption{Voltage as a function of time for both oscilloscope channels. The correspondence to the voltage indicated in figure~\ref{fig:char-circ} are given in the legend.}
  \label{fig:base}
\end{figure}
\begin{figure}
  \centering
  \includegraphics[width=.9\linewidth]{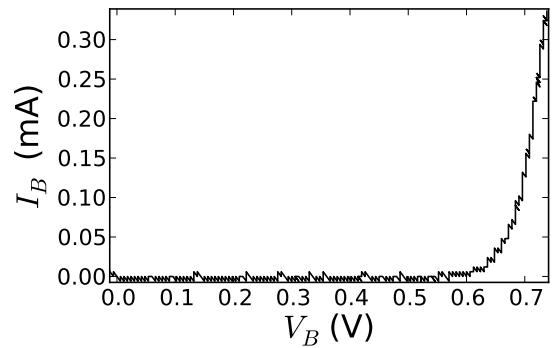}
  \caption{Base-Emitter characteristic.}
  \label{fig:BE}
\end{figure}

Figure~\ref{fig:base} displays the result the result obtained as it is visible on the oscilloscope, i.e. the voltage as a function of time for both channels. The instructions (aside for aesthetical details) to obtain Fig.~\ref{fig:base} are~:
\begin{verbatim}
>>> plot(CH1.t,CH1.V)
>>> plot(CH2.t,CH2.V)
\end{verbatim}
We then plot the current $I_B$ in function of $V_B$ in Fig.~\ref{fig:BE}. $I_B$ is computed from the oscilloscope measure as $\frac{V_1-V_B}{R_B}$. Assuming that channel~1 was plugged at the left of $R_B$ and channel~2 at the base, we display the characteristic~:
\begin{verbatim}
>>> RB=1000
>>> plot(CH2.V,(CH1.V-CH2.V)/RB)
\end{verbatim}

\subsection{The collector-emitter characteristic}
\label{sec:CE}

We make use of the same circuit as above to measure the collector-emitter characteristic with the difference that we now need to set $V_1$ to a suitable DC value and $V_2$ to a sawtooth signal of frequency $100$Hz in the range $0$V to $15.3$V. A period of the signal is needed to collect the data for one line of the characteristic (i.e. one value of $I_B$) and the procedure is repeated for different values of $I_B$. A voltmeter is used to measure the values of $V_B$ and $I_B$ as both oscilloscope channels are already occupied.

\begin{figure}
  \centering
  \includegraphics[width=.9\linewidth]{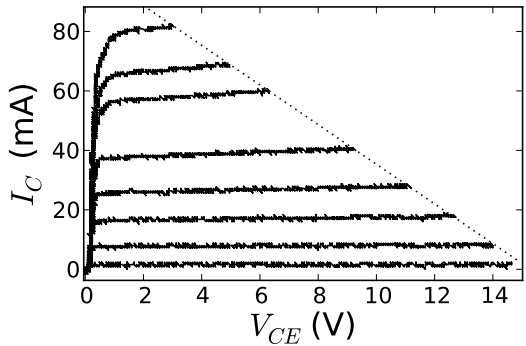}
  \caption{The Collector-Emitter characteristic curve. The dotted line corresponds to $I_C^{short}=\frac{V_{max} - V_{CE}}{R_C+R_{V_2}}$, the short circuit current for a given voltage ($R_{V_2}$ is the internal resistance of the voltage generator).}
  \label{fig:CE}
\end{figure}

The characteristic curve is $I_C$ as a function of $V_{CE}$. It is displayed in Fig.~\ref{fig:CE} making use of $I_C=(V_2-V_{CE})/R_C$ with the instructions~:
\begin{verbatim}
>>> RC=100
>>> plot(CH2.V,(CH1.V-CH2.V)/RC)
\end{verbatim}
assuming that channel~1 is plugged at the top of $R_C$ and channel~2 at the collector.
It should be noted that, unlike other apparatus,\cite{barnes_tpt_1974,camden_tpt_2005} the curves in Fig.~\ref{fig:CE} are not loops but go in one direction only.

The set of measure for different values of $I_B$ is collected in Fig.~\ref{fig:CE}. The low-$V_{CE}$ region displays a fast rise of the current (region of saturation) $I_C$ followed by a plateau (active region).
It is remarkable that the plateau becomes shorter as the current in the plateau increases. It is easily understood by drawing the characteristic curve of a power supply whose voltage is of $15.3$V and whose internal resistance is taken equal to $R_C+R_{V_2}$ where $R_{V_2}$ is the internal resistance of the voltage supply $V_2$. This characteristic curve defines the $V_{CE}$--$I_C$ relation in a polarized setting for the transistor.

\begin{figure}
  \centering
  \includegraphics[width=.9\linewidth]{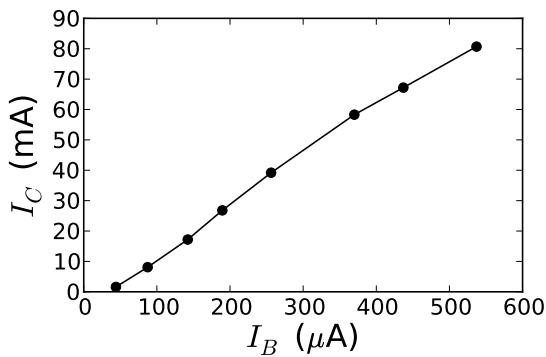}
  \caption{The collector current as a function of the base current. The slope of the line is found by linear regression and gives the gain $\beta = 164$. The base current is measured with a A-meter and the collector current is the value of the plateaus found in figure~\ref{fig:CE}.}
  \label{fig:gain}
\end{figure}

The average value of $I_C$ in the plateau is displayed as a function of $I_B$ in Fig.~\ref{fig:gain}. The linear behaviour illustrates the role of the transistor as a current amplifier. A linear fit gives a value of $\beta=164$ for the gain in current.


\section{Illustration with a common emitter circuit}
\label{sec:circuit}

\begin{figure}
  \centering
  \includegraphics[width=.9\linewidth]{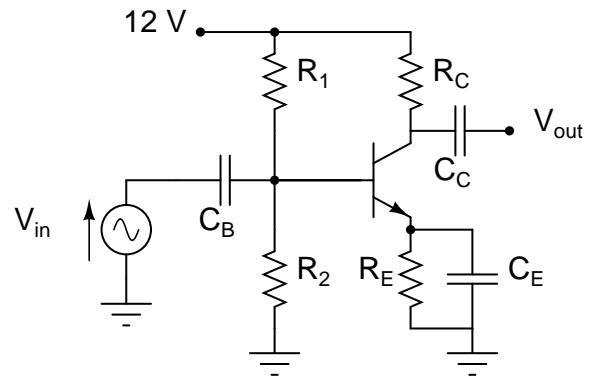}
  \caption{The common emitter circuit. $R_1=200$k$\Omega$, $R_2=40$k$\Omega$, $R_C=R_E=1$k$\Omega$ and all capacities are of $22\mu$F~.}
  \label{fig:com-emit}
\end{figure}

We present the measure of the voltage gain of a common emitter circuit.
The common emitter circuit is displayed in Fig.~\ref{fig:com-emit}, it is designed to operate at frequencies of about $1$kHz. The gain from this circuit is $G=\frac{V_{out}}{V_{in}}=-\beta R_C / r_\pi$ where $\beta$ is the current gain measured earlier, $R_C$ is the collector resistance and $r_\pi$ is the equivalent resistance in the hybrid-$\pi$ BJT model ($r_\pi=\beta r_e$ in Ref.~\onlinecite{malvino}) $r_\pi$ is the inverse of the slope of the base-emitter characteristic and its order of magnitude can be checked in Fig.~\ref{fig:BE}.
\begin{figure}
  \centering
  \includegraphics[width=.9\linewidth]{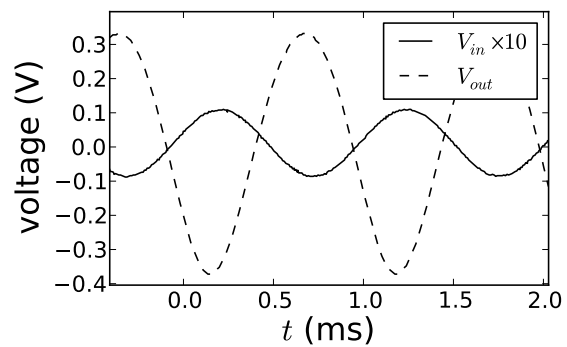}
  \caption{Measurement of $V_{in}$ and $V_{out}$ in the common emitter circuit of Fig.~\ref{fig:com-emit}. $V_{in}$ is magnified by $10$. The negative character of the gain is well displayed.}
  \label{fig:VinVout}
\end{figure}

A sinusoidal input signal of $10$mV of amplitude and $1$kHz of frequency is set on the function generator.
The measure of $V_{in}$ and $V_{out}$ is displayed in Fig.~\ref{fig:VinVout}, with
\begin{verbatim}
>>> plot(CH1.t,CH1.V)
>>> plot(CH2.t,CH2.V)
\end{verbatim}
Additionally, the DC measures at the pins of the transistor are~: $V_C=10.86$V, $V_B=1.738$V and $V_E=1.114$V. The polarization value for $V_{BE}$ is thus $0.624$V. The local slope of the curve $I_B(V_{BE})$ at the polarization point is equal to $1/r_\pi$.
The peak-to-peak voltage is the maximum value minus the minimum value of the signal. We compute the gain, up to the sign, via the instruction
\begin{verbatim}
>>> (CH2.V.max()-CH2.V.min())/
       (CH1.V.max()-CH1.V.min())
\end{verbatim}
and find $|G|\approx 34.9$~. Knowing the value of $\beta$ and of $R_C$, we deduce a value of $r_\pi\approx 4.7$k$\Omega$. This is in nice agreement with the inverse of the slope found in Fig.~\ref{fig:BE} around $V_{BE} = 0.624$V~.

As a final remark on the common-emitter circuit~: Fig.~\ref{fig:VinVout} displays a slight asymmetry that is caused by the limit of the linear approximation made in the modeling of the circuit. Indeed, the linear gain found between the input and output voltage supposes that $I_B(V_{BE})$ can be taken as a straight line. As soon as we deviate too much from the DC polarization, imperfections will occur.
An electronic course is a good introduction to the concept of modeling and to its limit, given its ability to check the theory and to see at which point it fails.

\section{Discussion}
\label{sec:disc}

We have presented in this article a simple setup allowing to take digital measurements on a transistor, giving the example of the measure of the characteristic curves. This setup represents an improvement in simplicity as compared to customized circuits found in the literature or with respect to commercial characteristic tracers.
The setup is of negligible cost, apart from the DSO, which is a quite common device nowadays. In addition, the use of open-source software allows the students to analyze data with no supplementary cost or to install the software freely in a computer lab.

From the pedagogical point of view, the setup we propose allows students to apprehend the physical working of a transistor with quantitative tools and observe themselves the limitations of the theory.
As mentioned in the introduction, the availability of a complete experimental process and apparatus, from building the experiment to interpreting the data, {\it via} the measurement, can help tailoring students' perception of modeling versus exact theory.

The speed with which characteristic curves are measured and imported on the computer makes it possible to include this experiment in place of the usual step by step measurement in the introductory lab of electronics. The time spent in explaining the technicalities is compensated by the short time needed to trace a full curve in the press of a button.
The generality of the setup, which allows to import easily ``real-world'' data onto the computer, will give it applications wherever an oscilloscope is useful but with a far more powerful analysis language. If desired by the teacher, the data can be re-used with stronger tools~: the computation of the frequency-response can be measured, or FFT's can be performed on the computer.

The author would like to acknowledge the helpful technical support from P. Duhamel and M. Trivilini of the physics department, support from Profs P. Gaspard and P. Emplit and from Dr N. Goldman, and the enthusiasm of students along the years.

\bibliographystyle{unsrt}

\begin{thebibliography}{10}

\bibitem{malvino}
A.P. Malvino and D.J. Bates.
\newblock {\em Electronic {P}rinciples}.
\newblock McGraw Hill, 7th edition, 2006.

\bibitem{gubanski_tpt_1973}
Zenon Gubanski.
\newblock Apparatus for teaching physics: Transistor characteristic curves on
  an oscilloscope.
\newblock {\em The Physics Teacher}, 11:359, Sep 1973.

\bibitem{barnes_tpt_1974}
George Barnes.
\newblock Apparatus for teaching physics: Families of transistor
  characteristics on an oscilloscope.
\newblock {\em The Physics Teacher}, 12:239--241, Apr 1974.

\bibitem{ponting_tpt_1991}
R.~Lee Ponting.
\newblock The oscilloscope as a primary measurement tool.
\newblock {\em The Physics Teacher}, 29:401--403, Sep 1991.

\bibitem{ramachandran_physed_1993}
V~Ramachandran.
\newblock Investigating the active region of transistor characteristics.
\newblock {\em Phys. Ed.}, 28:252--254, Jul 1993.

\bibitem{camden_tpt_2005}
M~Camden.
\newblock Oscilloscope display of current-voltage curves.
\newblock {\em The Physics Teacher}, 43:121, Feb 2005.

\bibitem{manzanares_ajp_1994}
Jos{\'e}~A Manzanares, Juan Bisquert, Germ{\`a} Garcia-Belmonte, and Mercedes
  Fern{\'a}ndez-Alonso.
\newblock An experiment on magnetic induction pulses.
\newblock {\em Am. J. Phys.}, 62:702--706, Aug 1994.

\bibitem{masters_ajp_1997}
M~F Masters and R~E Miers.
\newblock Use of a digital oscilloscope as a spectrum analyzer in the
  undergraduate laboratory.
\newblock {\em Am. J. Phys.}, 65:254--255, 1997.

\bibitem{potter_tpt_2003}
David Potter.
\newblock Phase changes in reflected sound waves.
\newblock {\em The Physics Teacher}, 41:12--13, Jan 2003.

\bibitem{martinez_ricci_ajp_2007}
M.~L~Mart{\'\i}nez Ricci, J~Mazzaferri, A.~V Bragas, and O.~E Mart{\'\i}nez.
\newblock Photon counting statistics using a digital oscilloscope.
\newblock {\em Am. J. Phys.}, 75:707--712, Aug 2007.

\bibitem{wadhwa_physed_2009}
Ajay Wadhwa.
\newblock Measuring the coefficient of restitution using a digital
  oscilloscope.
\newblock {\em Phys. Ed.}, 44:517--521, Sep 2009.

\bibitem{python}
Python programming language~: http://www.python.org/.

\bibitem{mpl}
matplotlib: Python plotting~: http://matplotlib.sourceforge.net/.

\end{thebibliography}

\end{document}